\documentclass[pre,onecolumn,showpacs,12pt]{revtex4}
\usepackage{amsmath,amsfonts}
\usepackage[dvips]{graphicx}
\usepackage{amsfonts}
\usepackage{amsmath}
\usepackage{indentfirst}
\usepackage{bm}
\usepackage{pstricks}
\usepackage{subfig}

\def\aa{\alpha}
\def\bb{\beta}
\def\GG{\Gamma}
\def\SDL{\Gamma_{\alpha,\beta}}
\def\disty{\displaystyle}

\begin{document}
\title{Complexity classification of quantum many-body systems
        according to the Pair of Order-Disorder Indices (PODI).}
\author{C.~P.~Panos}\email{chpanos@auth.gr}
\author{K.~Ch.~Chatzisavvas}\email{kchatz@auth.gr}
\affiliation{
Department of Theoretical Physics,\\ Aristotle University of Thessaloniki,\\
54124 Thessaloniki, Greece}
\date{\today}

\begin{abstract}
\noindent The \emph{statistical} measures of complexity $C(N)$
defined by L$\acute{\textrm{o}}$pez-Ruiz, Mancini, and Calbet
(LMC) and $\GG_{\aa,\bb}(N)$ according to Shiner, Davison and
Landsberg (SDL) are calculated as functions of the number of
particles $N$ for four quantum many-body systems, i.e. atoms,
nuclei, atomic clusters, and correlated atoms in a trap (bosons).
The strengths of \emph{disorder} $\aa$ and order $\bb$ are
evaluated for each system by imposing the condition
$\GG_{\aa,\bb}(N)=C(N)$. The proper pair $(\aa,\bb)$, obtained by
the above requirement, can serve as a Pair of Order-Disorder
Indices (PODI), characterizing quantitatively order versus
disorder in any quantum system. According to the above
classification, we assign to bosons the complexity character of
\emph{disorder}, to atoms the character of \emph{order}, while
nuclei and atomic clusters are (less) \emph{disordered} and lie
between them. This criterion can be used to estimate the relative
contribution of order and disorder to complexity for other more
complicated quantum systems as well and even classical ones, if
one is able to describe them probabilistically. We also address
the issue, whether those systems can grow in complexity as $N$
increases. Our comparative study indicates that atoms are the only
quantum system out of four, which is ordered, with the ability to
self-organize. We conjecture that this is an information-theoretic
reason that atoms are suitable as building blocks of larger
structures of biological interest i.e. molecules and
macromolecules. This is in the spirit of Wheeler's \emph{it from
bit} quote, the project to present everything about the Universe
in terms of information theory.
\end{abstract}
\pacs{\texttt{05.65.+b 02.50.-r}} \maketitle

\section{Introduction}\label{sec:Intro}

The question whether physical or biological systems can organize
themselves without the intervention of an external factor, is a
hot subject in the community of scientists interested in
complexity. A practical way to answer such a question is to use a
suitable definition of complexity and check if this quantity
increases with the number of particles $N$, implying the ability
or inability to self-organize. There are several measures of
complexity in the literature. One of them is the algorithmic
complexity according to Kolmogorov and Chaitin
\cite{Kolmogorov65,Chaitin66} defined as the length of the
shortest possible program necessary to reproduce a given object.
The fact that a given program is indeed the shortest one, is hard
to prove. In this paper we employ the \emph{statistical} measure
of complexity $C$, defined by L$\acute{\textrm{o}}$pez-Ruiz,
Mancini, and Calbet (LMC) \cite{Lopez95}, which can be calculated
easily, provided that the information content of a quantum system
is known from previous work \cite{SearsPHD}-\cite{PaperPCMK}. In
the same spirit, we use the measure of complexity $\SDL$ according
to Shiner, Davison and Landsberg (SDL) \cite{Shiner99}.

We intend to calculate and compare $C(N)$ with $\SDL(N)$ as
functions of the number of particles $N$ in four quantum systems,
namely atomic nuclei, atoms, atomic clusters, and correlated atoms
in a trap--bosons. Here, we use continuous probability density
distributions of particles of quantum systems, while in
\cite{Panos08} we obtained $(\aa,\bb)$ using $C(Z)$ and
$\GG_{\aa,\bb}(Z)$, taking into account discrete probability
distributions of electron configurations in atoms.

It is noted that such complexity measures, based in a
probabilistic description of a quantum system, have been
calculated quantitatively (as functions of $Z$) for the first time
in \cite{PaperCMP,PaperPCMK}, using atoms as a test bed. A
previous preliminary step was the finding that Landsberg's order
parameter \cite{Landsberg84} is an increasing function of $N$ for
fermions and bosonic systems \cite{PaperMMP} (see also
\cite{Panos01}).

The aim of the present work is to find specific overall values of
$(\aa,\bb)$ for each system, by requiring that approximately
$\GG_{\aa,\bb}(N)=C(N)$. We investigate the possibility whether
the pair $(\aa,\bb)$, chosen in such a way (PODI: Pair of
Order-Disorder Indices) is useful for the classification of
quantum many-body systems according to the contribution of order
and disorder to complexity.

Our paper is organized as follows: In Sec. \ref{sec:Measures} we
define measures of information and complexity, in Sec.
\ref{sec:Numerical} we present our numerical results and Sec.
\ref{sec:Conclusions} contains our conclusions.

\section{Measures of information content and complexity of a
system}\label{sec:Measures}

Shannon's information entropy \cite{Shannon48}, is defined as
\begin{eqnarray}
    S_r&=&-\int
    \rho(\textbf{r})\,\ln{\rho(\textbf{r})}\,d\textbf{r} \label{eq:Sr}\\
    S_k&=&-\int n(\textbf{k})\,\ln{n(\textbf{k})}\,d\textbf{k} \label{eq:Sk}
\end{eqnarray}
where $\rho(\textbf{r})$ and $n(\textbf{k})$  are normalized to
one density distributions in position and momentum spaces
respectively. $S_{r}$, $S_{k}$ depend on the units used to measure
$\textbf{r}$ and $\textbf{k}$, while the important quantity is the
sum $S=S_{r}+S_{k}$, which is scale invariant i.e. independent of
units. There is a delicate balance between conjugate spaces
leading to the following results:

First, the entropic uncertainty relation (EUR) holds of the form
$S=S_{r}+S_{k}\geq 3(1+\ln{\pi})$ for a 3-dimensional quantum
system \cite{Bialynicki75}. The lower bound is attained for a
Gaussian distribution. The above inequality is stronger than the
Heisenberg uncertainty relation, because the right-hand side of
EUR does not depend on the state, while Heisenberg's does depend.
Also Heisenberg's inequality can be derived from EUR, while the
inverse is not true. $S$ represents the information content of the
system in bits, if one uses logarithm with base 2 and in nats
(natural unit of information), if the logarithm is natural.

Second, the universal property $S=a+b\,\ln{N}$ was proposed and
verified for various quantum systems (nuclei, atoms, atomic
clusters and correlated atoms in a trap) \cite{PaperMP2},
\cite{PaperMP}. The parameters $a$, $b$ depend on the system under
consideration. That property holds for systems of various sizes,
obeying different statistics (fermions or bosons) with different
numbers of particles and various interactions. Specifically, the
sizes of those systems range from the order of fermis ($10^{-13}$
cm) in nuclei, to $10^4$ $\AA$ ($10^{-4}$ cm) for bosonic systems,
while the number of particles $N$ starts from a few to a hundred
and goes up to millions (bosons).

The Shiner, Davison and Landsberg (SDL) measure complexity
$\GG_{\aa,\bb}$ is defined as:
\begin{equation}\label{eq:gamma}
    \GG_{\aa,\bb}=\Delta^{\aa} \cdot \Omega^{\bb}=
    \Delta^{\aa}(1-\Delta)^{\bb}%=\Omega^{\bb}(1-\Omega)^{\aa}
\end{equation}
where
\begin{equation}\label{eq:delta}
    \Delta=\frac{S}{S_{\rm max}}\quad \mbox{and} \quad \Omega=1-\Delta
\end{equation}
are the normalized measures of disorder and order respectively,
according to Landsberg \cite{Landsberg84}, obeying the
inequalities $0<\Delta <1, 0<\Omega <1$.

The parameter $\aa$ represents the strength of disorder , while
$\bb$ the strength of order. We have a complexity measure
$\GG_{\aa,\bb}$ of category I, if $\bb=0$ and $\aa >0$, where
complexity is an increasing function of disorder. In category II
($\aa >0,\bb >0$) $\GG_{\aa,\bb}$ is a convex function, while in
category III ($\aa=0,\bb >0$)  $\GG_{\aa,\bb}$ is a decreasing
function of disorder $\Delta$.

The LMC measure $C$ \cite{Lopez95} is defined as
\begin{equation}\label{eq:eq1}
    C=S\cdot D
\end{equation}
where $S$ denotes the information content stored in the system (in
our case Shannon's information entropy sum $S=S_{r}+S_{k}$
\cite{Shannon48})
 and $D$ is the disequilibrium of the system.

For a discrete probability distribution $\{p_{i}\}$, the
disequilibrium $D$ can be defined as the quadratic distance of
$\{p_{i}\}$ to the equiprobable (uniform) state
$\disty{p_{i}=\frac{1}{n}}, i=1,2,\ldots$. Thus
\[
    D=\sum_{i=1}^{k} \left(p_{i}-\frac{1}{n} \right)^2
\]
while for a continuous probability distribution $\rho(\bf{r})$, it
is extended in position-space as
\[
    D_r=\int \rho^2(\textbf{r}) \,d(\textbf{r})
\]
and in momentum-space
\[
    D_k=\int n^2(\textbf{k}) \,d(\textbf{k})
\]
The extension from the discrete to the continuous case is
justified in \cite{PaperMMP}. The proper combination of $D_r$ and
$D_k$, to be inserted into (1) is
\begin{align}
    D=D_r \cdot D_k
\end{align}
In fact $D_r$ has the dimension of inverse volume, while $D_k$ of
volume, leading to the dimensionless quantity $D$.

\section{Numerical Results}\label{sec:Numerical}

We intend to calculate numerically the LMC measure of complexity
$C(N)=S\cdot D$ and the SDL one, $\GG_{\aa,\bb}(N)=\Delta^{\aa}
\cdot \Omega^{\bb}$ as functions of the number of particles $N$
(electrons in atoms, nucleons in nuclei, valence electrons in
atomic clusters, and alkali atoms in bosonic traps). Our
calculations are facilitated by our previous research in
information entropy $S$. The key quantities of our calculations
are the density distributions in position-space $\rho(\textbf{r})$
and momentum-space $n(\textbf{k})$, obtained as follows.

For nuclei we performed microscopic mean-field calculations
\cite{PaperMP2}, using a density-dependent Skyrme force, in
particular the SKIII force \cite{Dover72}. For atoms we used the
RHF (Roothaan-Hartree-Fock) electron distributions \cite{Bunge93},
applied to calculations of atomic complexity \cite{PaperCMP},
\cite{PaperPCMK}. For atomic clusters, we employed the jellium
model (Ekcard model) \cite{PaperMP2}, \cite{Ekardt84},
\cite{Nishioka90}, while for bosons in a trap we carried out a
numerical calculation of densities in both position- and
momentum-spaces, solving numerically the non-linear
Gross-Pitaevski equation \cite{PaperMMP}. In all of the above
calculations special care has been devoted to an accurate
treatment of the Fourier transform of $\rho(\textbf{r})$ in order
to obtain $n(\textbf{k})$ for each system.

Next, the distributions $\rho(\textbf{r})$ and $n(\textbf{k})$
calculated as described above, are inserted into relations
(\ref{eq:Sr}) and (\ref{eq:Sk}) giving $S_r(N)$ and $S_k(N)$ for
each system, needed to find the sum $S(N)=S_r(N)+S_k(N)$. Thus,
$C(N)=S(N)\cdot D(N)$ can be obtained easily. However, in order to
evaluate $\GG_{\aa,\bb}(N)=\Delta(N)^{\aa} \cdot
(1-\Delta(N))^{\bb}$, we need in addition $\Delta(N)=S(N)/S_{\rm
max}(N)$, where $S_{\rm
max}(N)=S_{r\textrm{max}}(N)+S_{k\textrm{max}}(N)$.

For discrete probability distributions the value of $S_{\rm max}$,
is attained for an equiprobable (uniform) density distribution
i.e. $S_{\rm max}=\ln{N}$, where $N$ is the number of states with
probabilities $\{p_i\}, i=1,2,\ldots N$. $S$ is minimum, $S_{\rm
min}=0$, when only one probability is different from 0,
specifically $p_i=1$ for a fixed  $i$, while all the others
vanish. For continuous probability distributions, calculations for
$S_{\rm min}$ and $S_{\rm max}$ are based to the following
inequalities, introduced for the first time and verified for atoms
by Gadre and collaborators \cite{GadreSearsEtAlPaper},
\cite{SenEdBook}, for atomic clusters and nuclei in \cite{PaperMP}
and bosonic systems (correlated alkali atoms in a trap) in
\cite{PaperMMP} (see also \cite{PaperMMPNova}).

\begin{eqnarray}
    S_{r\textrm{min}}\leq & S_r & \leq  S_{r\textrm{max}} \label{eq:ineqSr} \\
    S_{k\textrm{min}} \leq & S_k & \leq S_{k\textrm{max}} \label{eq:ineqSk} \\
    S_{\textrm{min}} \leq & S & \leq  S_{\textrm{max}} \label{eq:ineqS}
\end{eqnarray}

where $S=S_r+S_k$ and $\rho(\textbf{r})$, $n(\textbf{k})$ are
normalized to 1, i.e. $\displaystyle{\int \rho(\textbf{r})\,
d\textbf{r}=1}$ and $\displaystyle{\int n(\textbf{k})\,
d\textbf{k}=1}$.

The lower and upper limits for density distributions normalized to
one, are:
\begin{align}
    S_{r\textrm{min}} &= \frac{3}{2}\, (1+\ln{\pi})-
    \frac{3}{2}\, \ln{(\frac{4}{3}T)} \nonumber \\
    S_{r\textrm{max}} &= \frac{3}{2}\, (1+\ln{\pi})+
    \frac{3}{2}\, \ln{(\frac{2}{3}\langle r^2 \rangle)}
\end{align}
and
\begin{align}
    S_{k\textrm{min}} &= \frac{3}{2}\, (1+\ln{\pi})-
    \frac{3}{2}\, \ln{(\frac{2}{3}\langle r^2 \rangle)} \nonumber \\
    S_{r\textrm{max}} &= \frac{3}{2}\, (1+\ln{\pi})+
    \frac{3}{2}\, \ln{(\frac{4}{3}T)}
\end{align}
Thus
\begin{align}
    S_{\textrm{min}} &= 3 \, (1+\ln{\pi}) -
    \frac{3}{2}\, \ln{\left(\frac{8}{9}\langle r^2 \rangle T
    \right)}
    \\
    S_{\textrm{max}} &=  3 \, (1+\ln{\pi})+
    \frac{3}{2}\, \ln{\left(\frac{8}{9}\langle r^2 \rangle T
    \right)} \label{eq:Smax}
\end{align}
where $\langle r^2 \rangle$ is the mean square radius of the
system and $T$ its kinetic energy. We use, as an input in
$\Delta=\frac{S}{S_{\rm max}}$, the values of $S_{\rm max}$
according to relation (\ref{eq:Smax}).

We repeated and verified the above inequalities for atoms in
\cite{PaperCMP}, \cite{PaperPCMK} and next, we checked their
validity numerically for nuclei, atomic clusters and bosonic traps
in \cite{PaperMMP}, \cite{PaperMP2}. We note that $\langle r^2
\rangle$ and $T$, can be calculated employing $\rho(\textbf{r})$
and $n(\textbf{k})$. Thus, we obtain the function $S_{\rm max}(N)$
for each quantum system.

\begin{figure}[h]
 \centering
    \includegraphics[height=7.0cm,width=7.0cm]{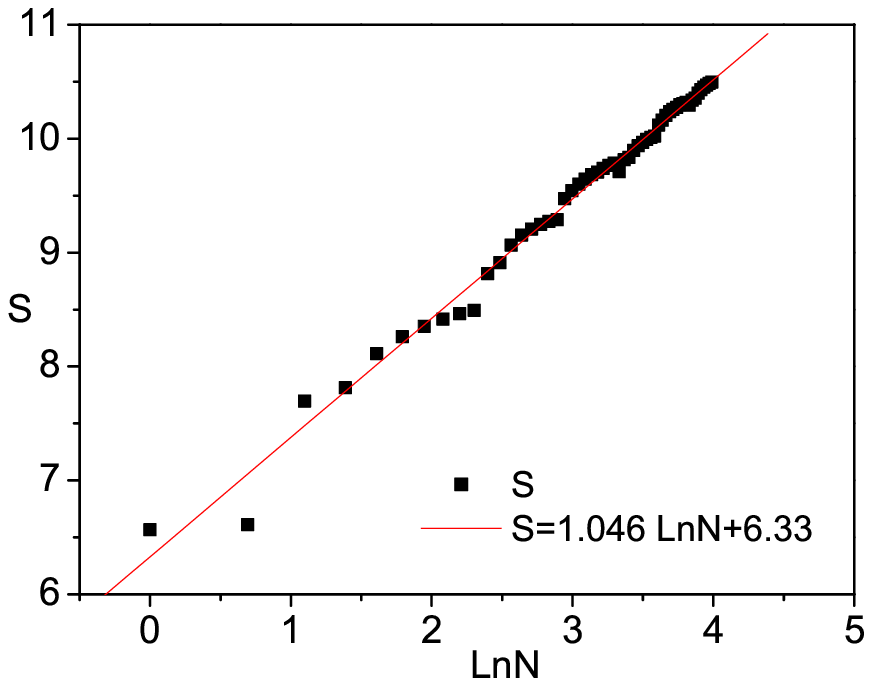}
    \includegraphics[height=7.0cm,width=7.0cm]{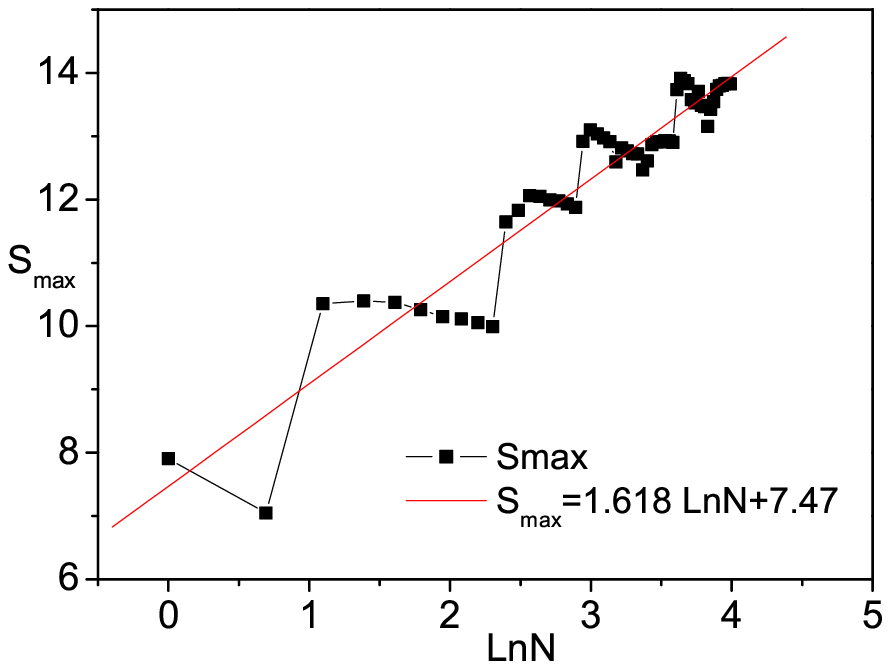}
    \\
    \includegraphics[height=7.0cm,width=7.0cm]{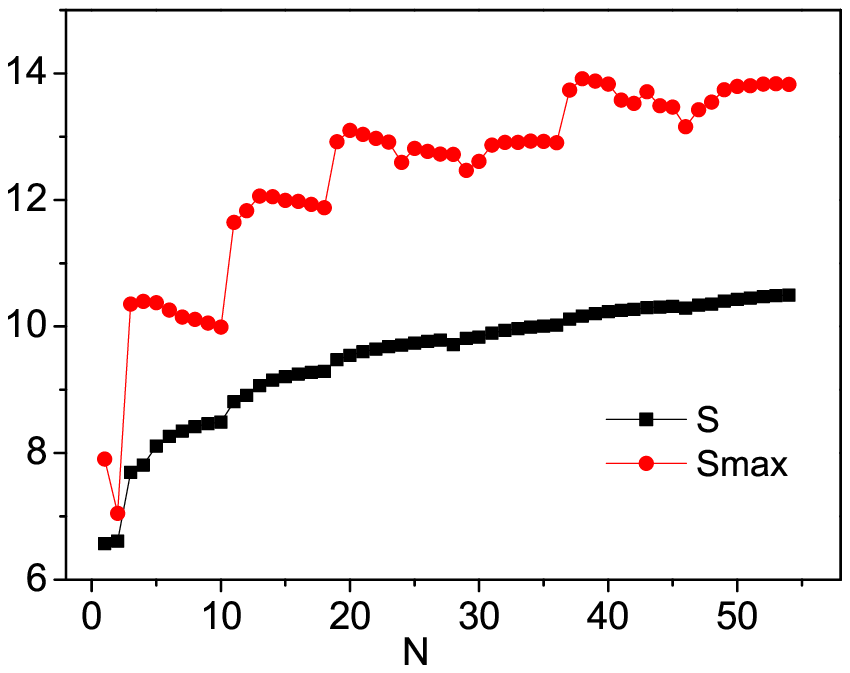}
    \includegraphics[height=7.0cm,width=7.0cm]{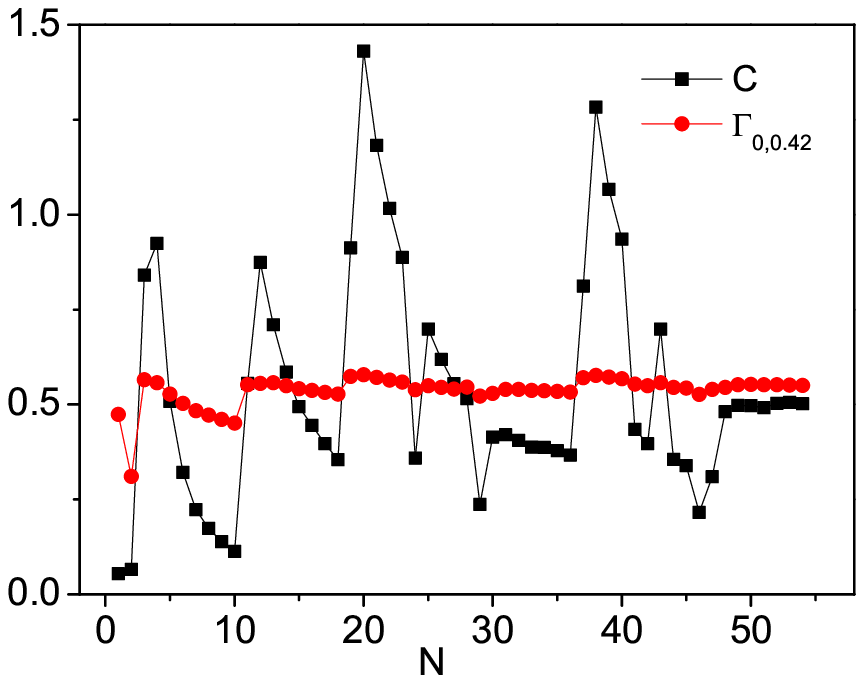}
    \\
    \includegraphics[height=7.0cm,width=7.0cm]{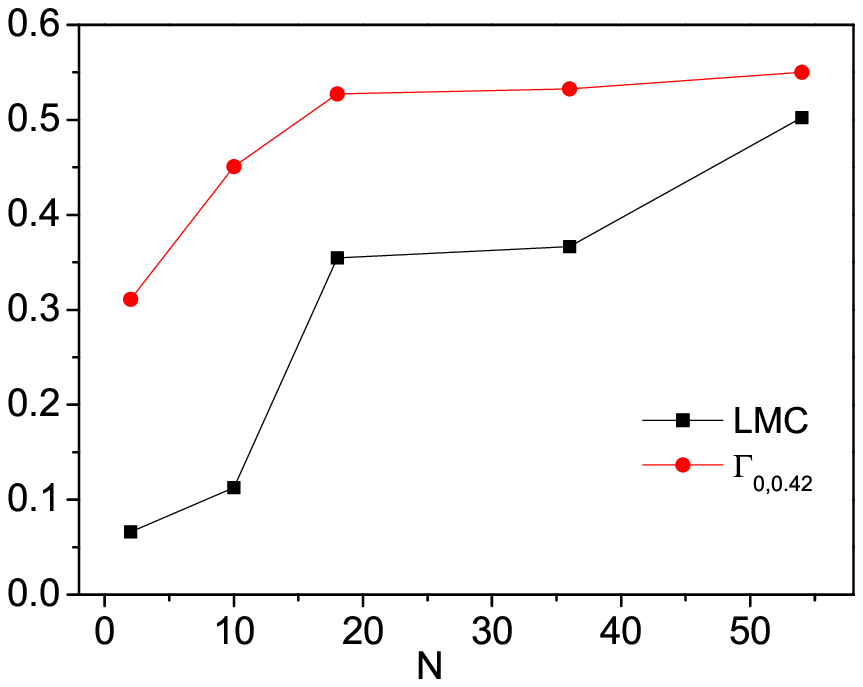}
    \includegraphics[height=7.0cm,width=7.0cm]{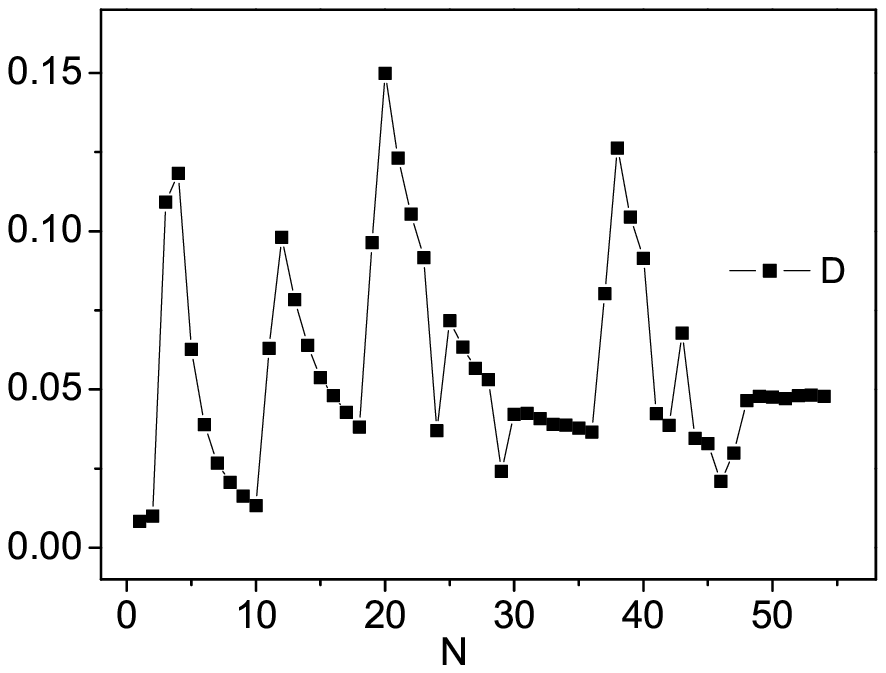}
    \caption{Atoms}\label{fig:resatoms}
\end{figure}

\begin{figure}[h]
 \centering
    \includegraphics[height=7.0cm,width=7.0cm]{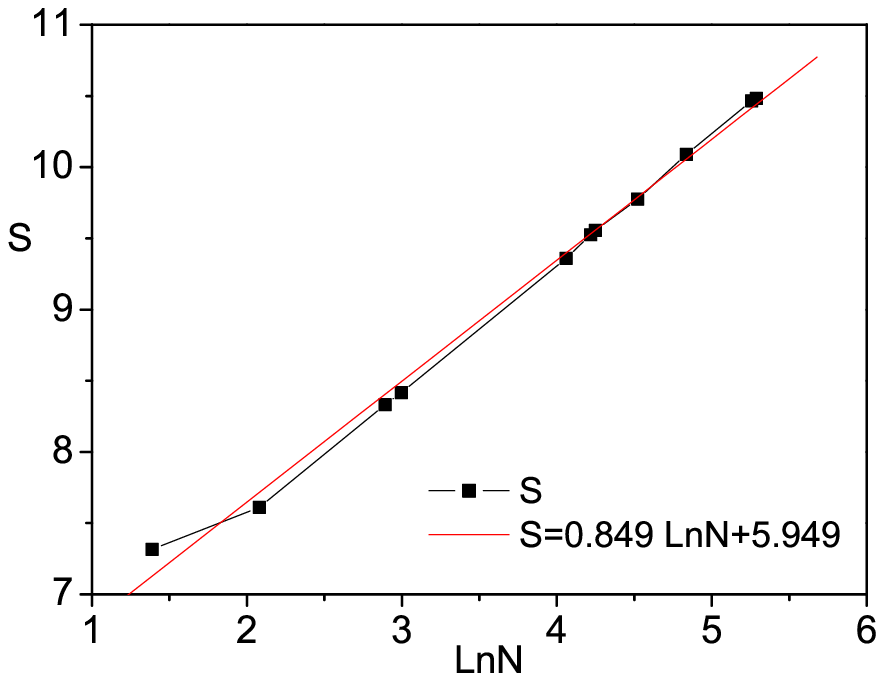}
    \includegraphics[height=7.0cm,width=7.0cm]{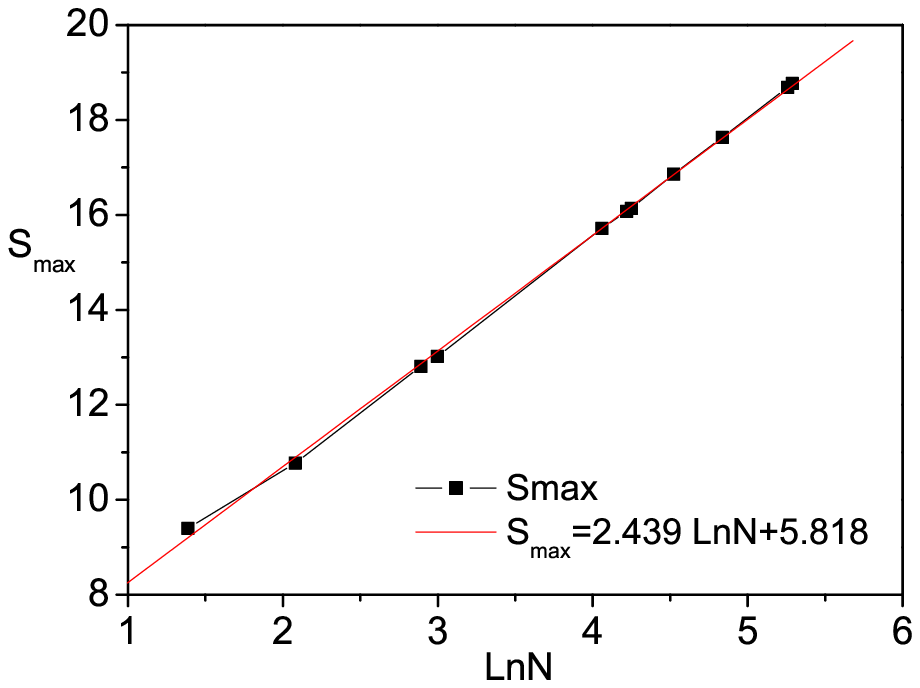}
    \\
    \includegraphics[height=7.0cm,width=7.0cm]{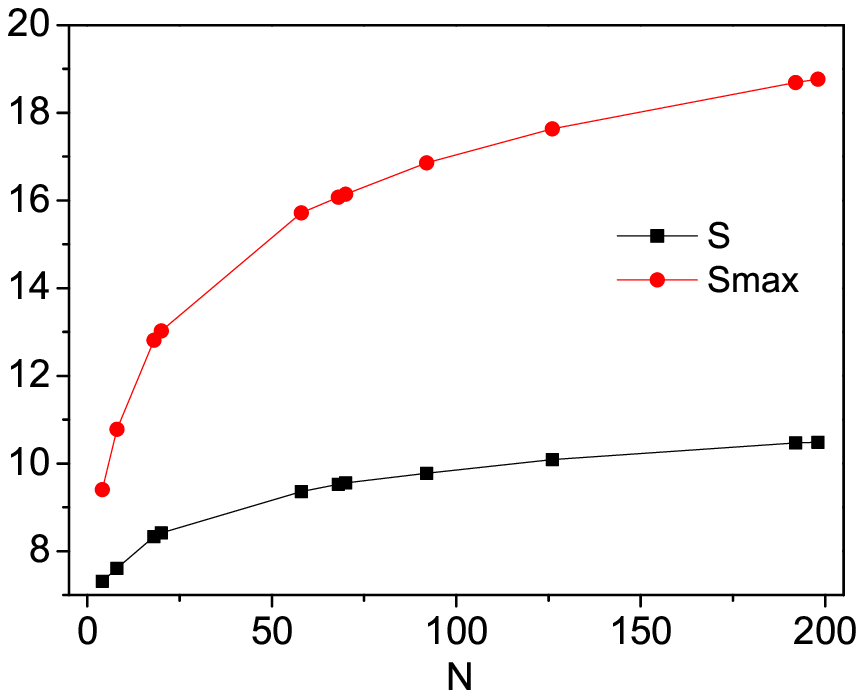}
    \includegraphics[height=7.0cm,width=7.0cm]{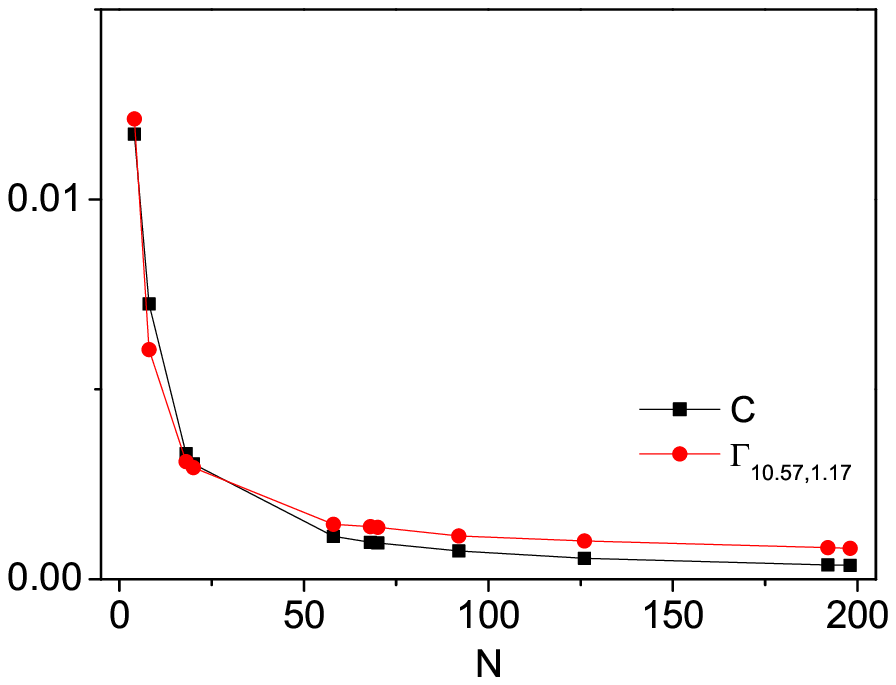}
    \\
    \includegraphics[height=7.0cm,width=7.0cm]{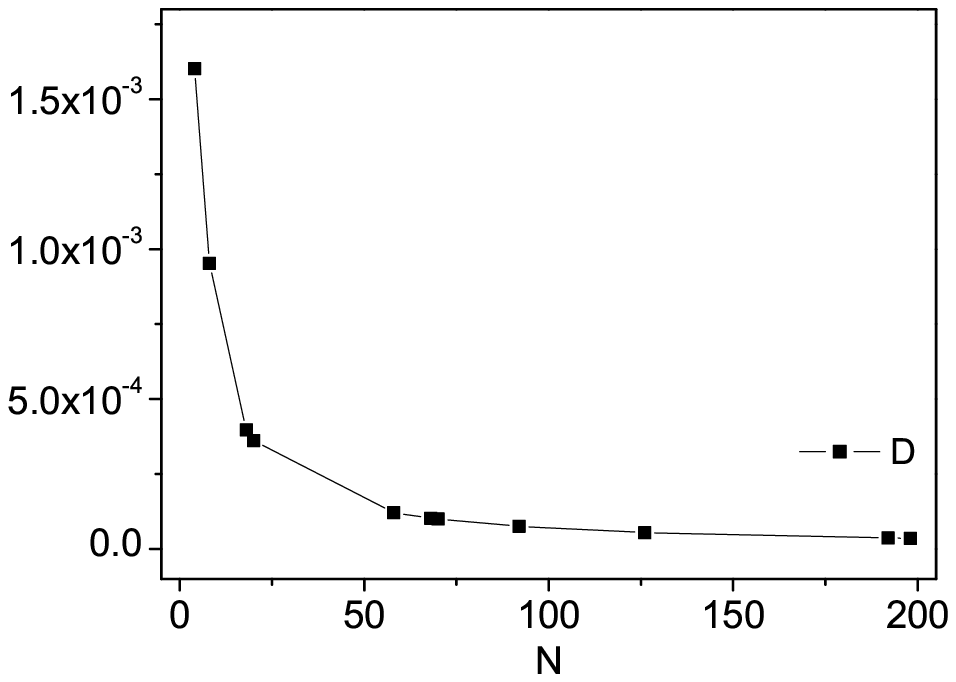}
    \caption{Atomic Clusters}\label{fig:rescluster}
\end{figure}

\begin{figure}[h]
 \centering
    \includegraphics[height=7.0cm,width=7.0cm]{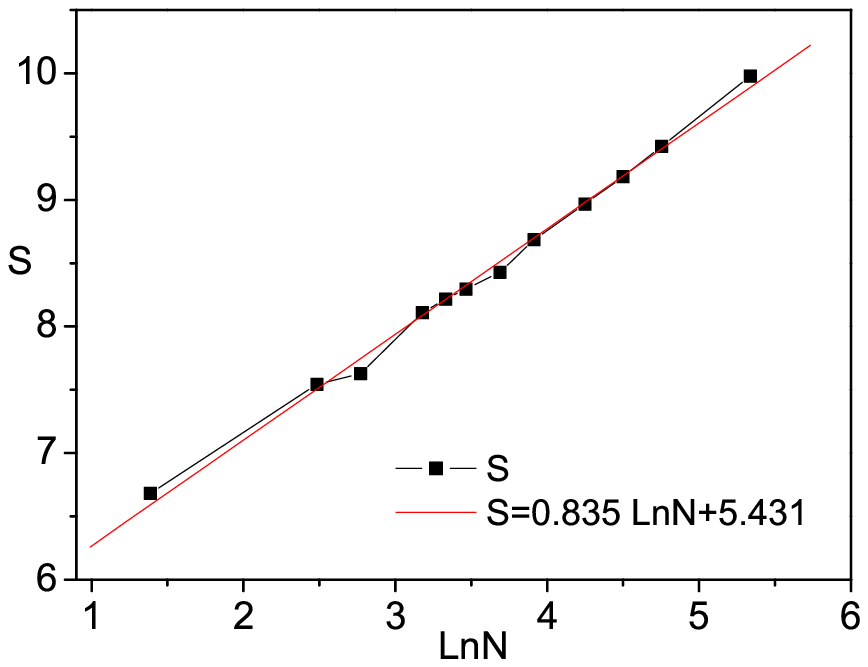}
    \includegraphics[height=7.0cm,width=7.0cm]{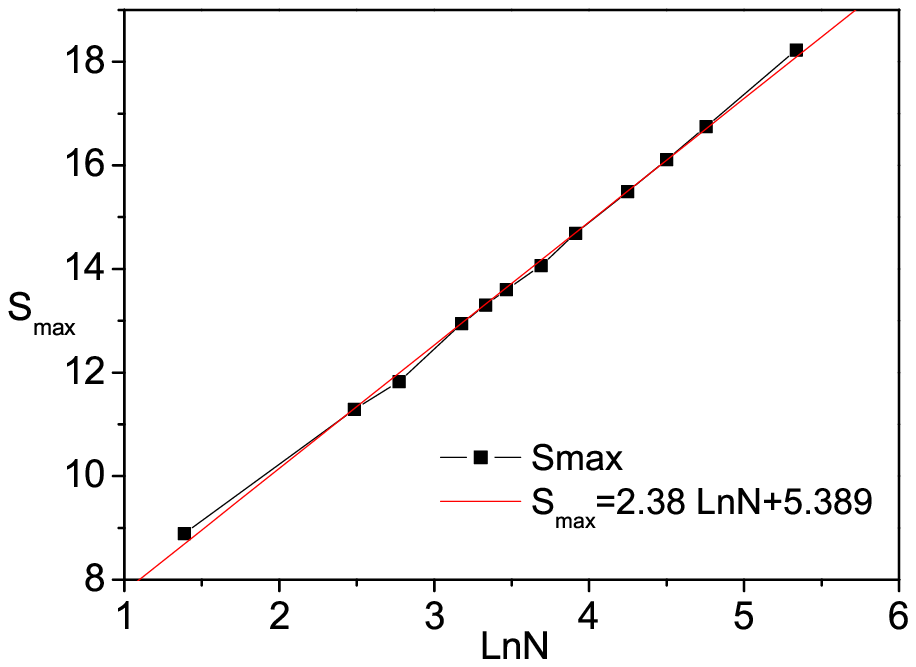}
    \\
    \includegraphics[height=7.0cm,width=7.0cm]{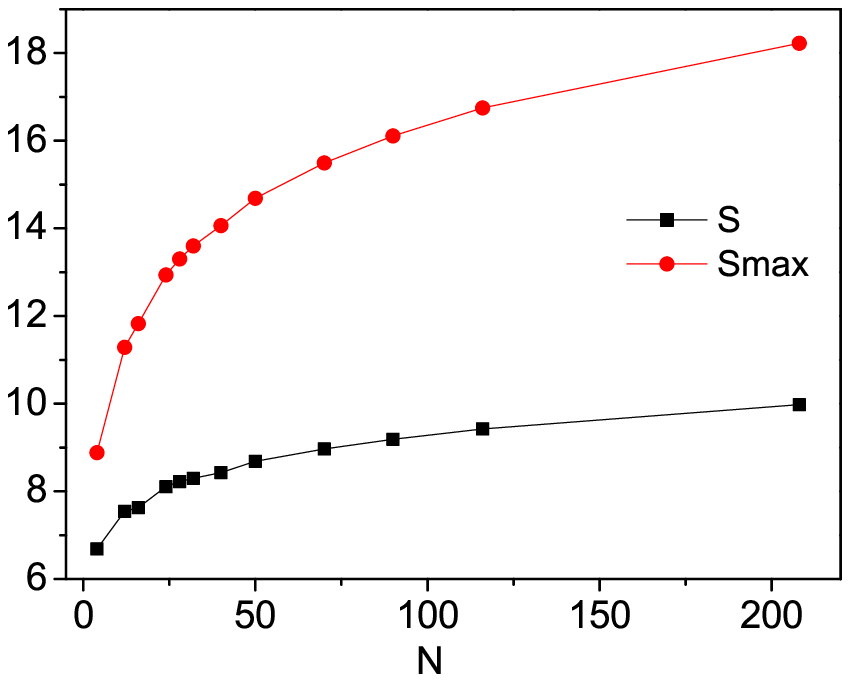}
    \includegraphics[height=7.0cm,width=7.0cm]{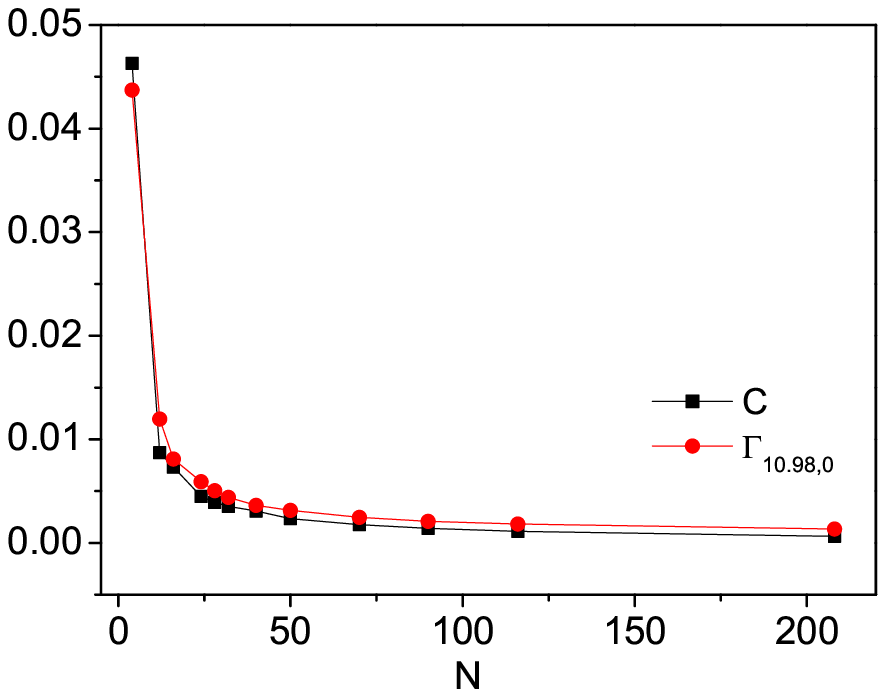}
    \\
    \includegraphics[height=7.0cm,width=7.0cm]{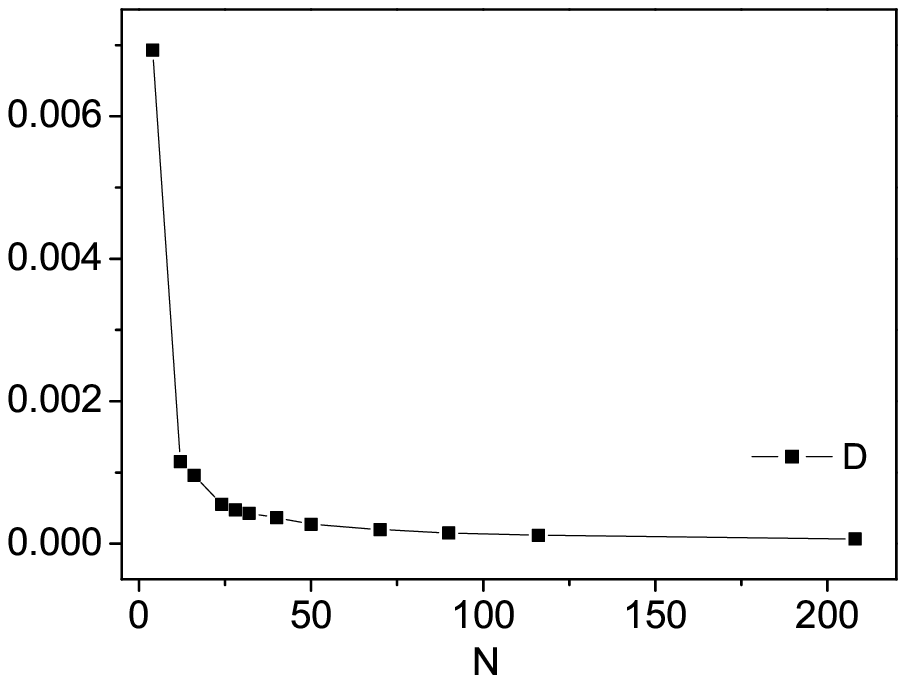}
    \caption{Nuclei}\label{fig:resnuclei}
\end{figure}

\begin{figure}[h]
 \centering
    \includegraphics[height=7.0cm,width=7.0cm]{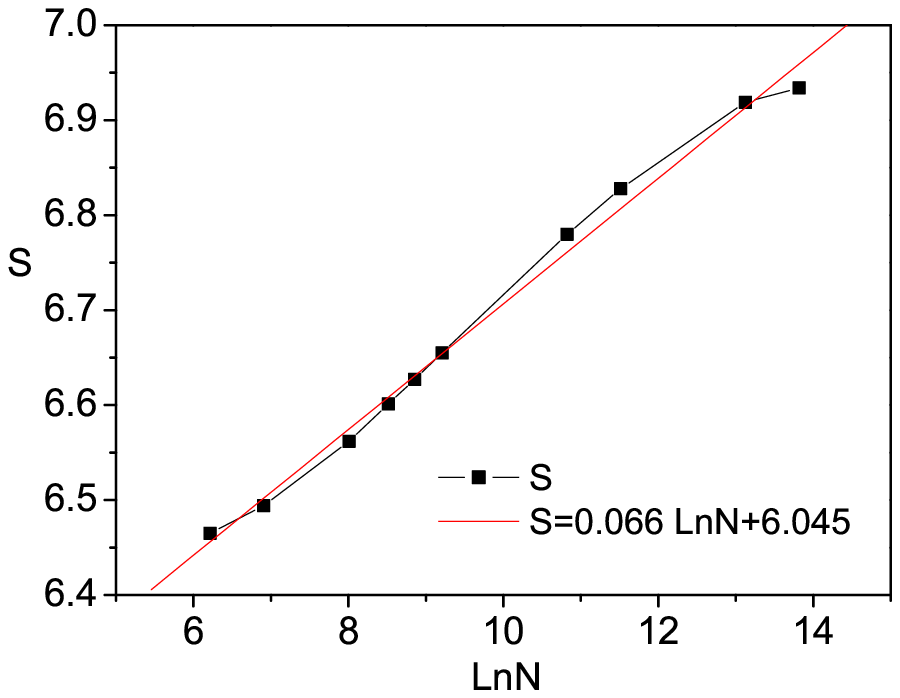}
    \includegraphics[height=6.0cm,width=7.0cm]{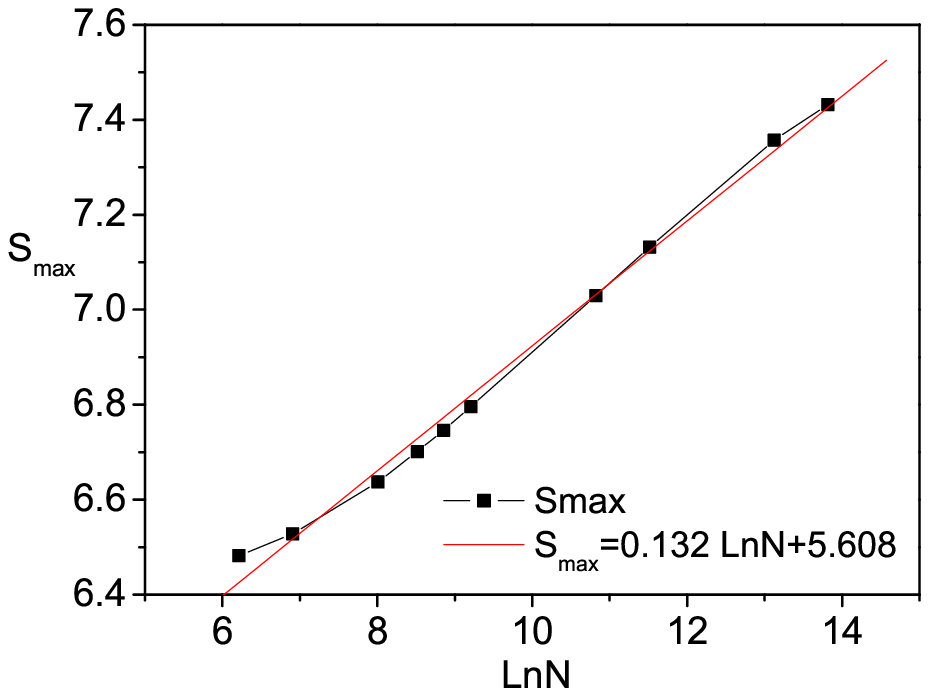}
    \\
    \includegraphics[height=7.0cm,width=7.0cm]{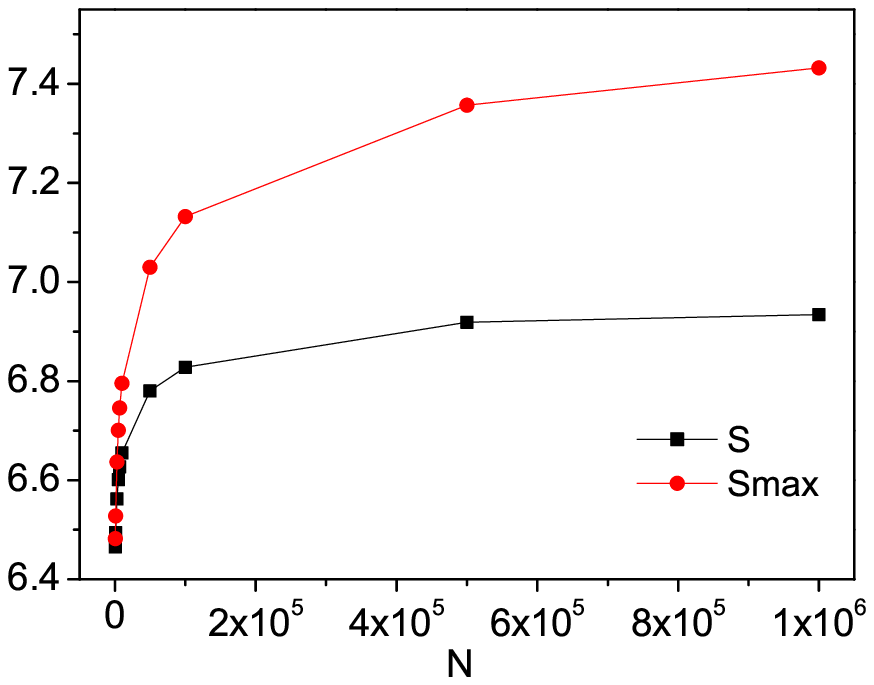}
    \includegraphics[height=7.0cm,width=7.0cm]{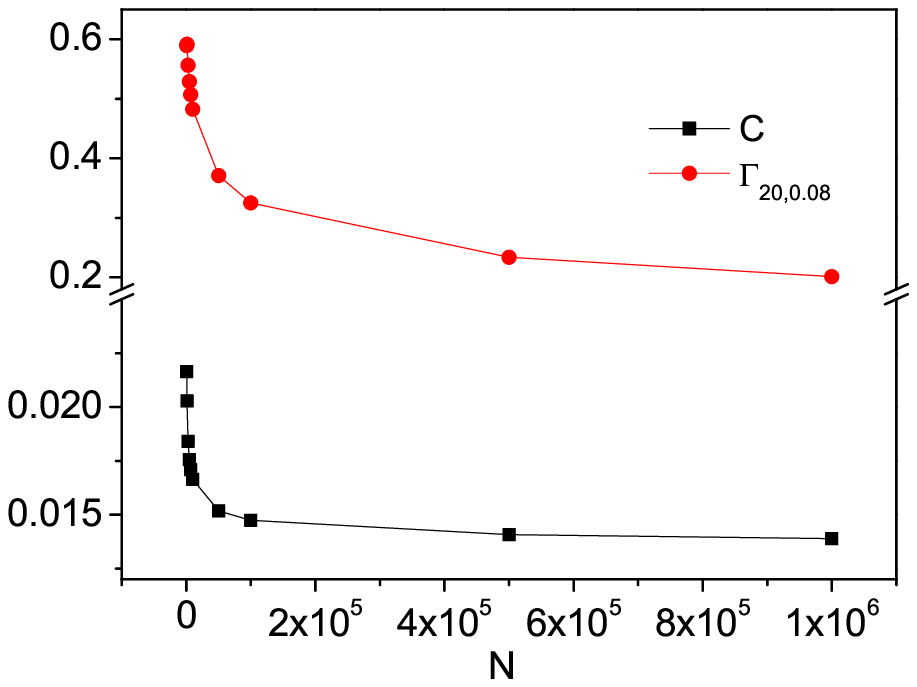}
    \\
    \includegraphics[height=7.0cm,width=7.0cm]{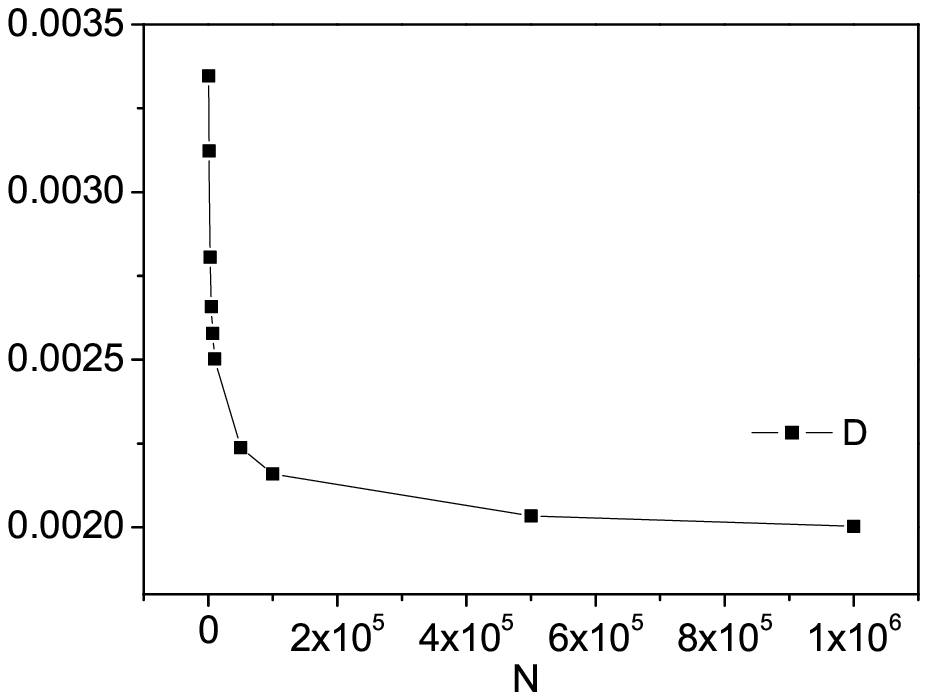}
    \caption{Bosons}\label{fig:resboson}
\end{figure}

Following the procedure described above, we calculate and plot the
functions $S(N)$, $S_{\rm max}(N)$, $C(N)$ and $\GG_{\aa,\bb}(N)$
(for various values $(\aa,\bb)$), and $D(N)$, for all four quantum
systems under consideration. Our results are presented for atoms
in Fig. \ref{fig:resatoms}, atomic clusters in Fig.
\ref{fig:rescluster}, nuclei in Fig. \ref{fig:resnuclei} and
bosons in a trap in Fig. \ref{fig:resboson}. We note that in Fig.
\ref{fig:resatoms} we present both the dependence of $C$ on all
values of $N$ and separately the dependence of $C$ only on values
of $N$ for closed shells atoms. The data of the latter figure is
used in the fitting of $\SDL (N)$ to $C(N)$. We also plot linear
fitted expressions for $S$ and $S_{\rm max}$ of the form
$S=a+b\,\ln N$. In order to choose the specific values
($\aa,\bb$), we employ a new prescription, described below.

We begin with the requirement that the curves  $\GG_{\aa,\bb}(N)$
and $C(N)$ should coincide (approximately) or show the same
pattern, for a proper pair of values $(\aa,\bb)$. We quantify the
similarity between the two curves with a norm
$\displaystyle{\sum_{i=1}\left( C_i - (\GG_{\aa,\bb})_i
\right)^2}$, which should be zero in the ideal case, where the two
curves are exactly the same. A comparison of $\SDL (N)$ with
$C(N)$ has been suggested for the first time in \cite{PaperPCMK},
where we obtained by inspection of the figures that roughly $\aa
\simeq 0$ and $\bb \simeq 4$ ($\aa <\bb$) and has been observed
that complexity increases with $N=Z$, based on the trend of closed
shells.

We limit our search for the optimal pair  $(\aa,\bb)$ only for the
regions of the $\aa-\bb$ plane, where the behavior of the
complexity measure $\GG_{\aa,\bb}$ is the same with the
corresponding behavior of $C(N)$.  Thus, if for example, $C(N)$ is
a decreasing function of $N$, then we search for the proper pair
$(\aa,\bb)$ only in the region where $\GG_{\aa,\bb}$ is decreasing
with $N$. The trend of complexity is dictated by $C(N)$.

The resulting pair of values $(\aa,\bb)$ is named \emph{Pair of
Order-Disorder Indices (PODI)}. We hope that it will serve as a
useful tool in order to classify quantum (or classical) many body
systems, according to the contribution of order versus disorder
contributions to complexity.

\begin{figure}[ht]
\centering
    \subfloat[Atoms]{
    %\label{fig:subfig:a} %% label for first subfigure
    \includegraphics[height=7.0cm,width=7.0cm]{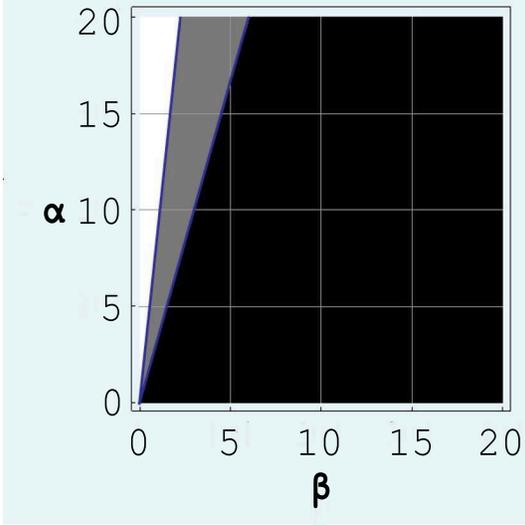}}
    \hspace{1cm}
    \subfloat[Atomic Clusters]{
    %\label{fig:subfig:b} %% label for second subfigure
    \includegraphics[height=7.0cm,width=7.0cm]{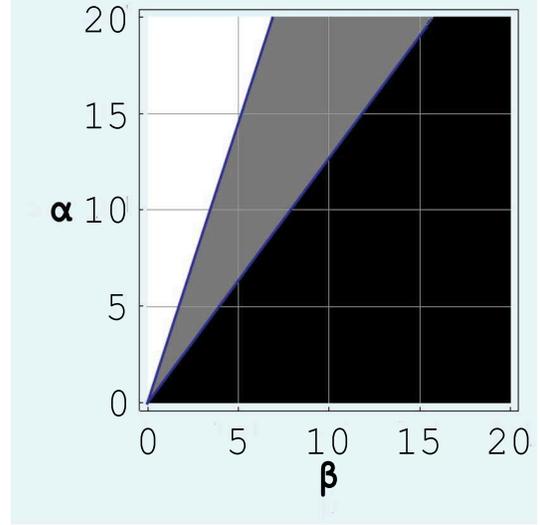}}
    \\
    %\label{fig:subfig:c} %% label for first subfigure
    \subfloat[Nuclei]{
    \includegraphics[height=7.0cm,width=7.0cm]{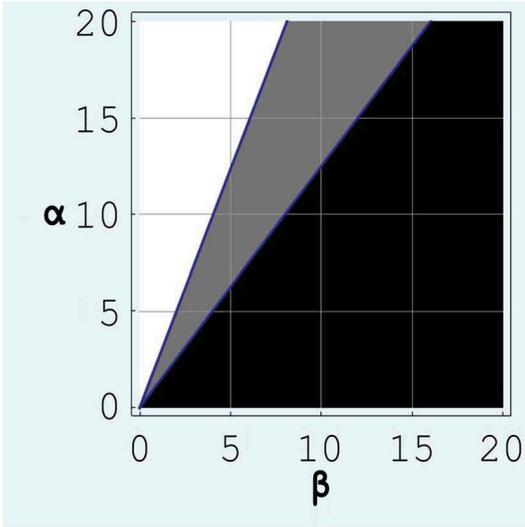}}
    \hspace{1cm} \subfloat[Bosons]{
    %\label{fig:subfig:d} %% label for second subfigure
    \includegraphics[height=7.0cm,width=7.0cm]{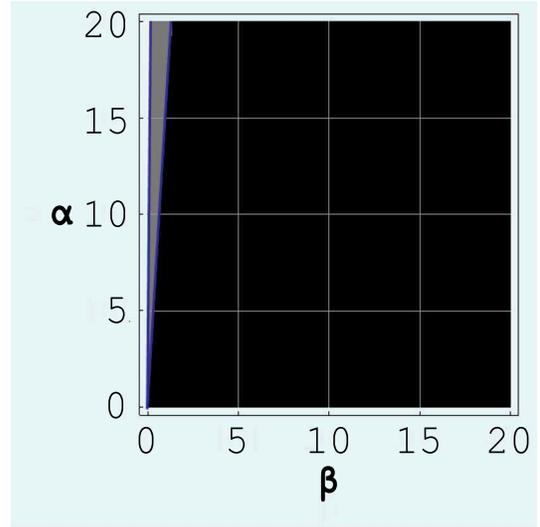}}
    \caption{Complexity regions in the $\aa-\bb$ plane,
    showing a decreasing trend of $\GG_{\aa,\bb}(N)$ (white), a convex (grey) and
    an increasing one (black), for (a) atoms, (b) atomic clusters, (c) atomic nuclei, and (d) bosons.
    The approximate linear boundaries of the three regions are given
    in Table \ref{tab:table}.} \label{fig:complexityclass}
\end{figure}

In Fig. \ref{fig:complexityclass}, we present our results for all
four quantum systems under consideration in the $\aa-\bb$ plane
with $(0\leq \aa \leq 20$, and $0\leq \bb \leq 20)$, with mesh
points $10^{-2}$ for $\aa$ and $\bb$. A decreasing behavior is
denoted by white color, a convex behavior by grey and an
increasing trend by black.

Qualitatively, the structure of the corresponding plots in
$\aa-\bb$ plane and the shape of the complexity regions are the
same for all considered quantum systems. A monotonically
decreasing region is followed by a convex one and then we have a
monotonically increasing area.

Our results are summarized in Table \ref{tab:table}, where in
addition we present the approximate linear limits of the three
respective complexity regions (decreasing, convex, increasing).

\begin{table}[h]
\begin{tabular}{|c|ccc|cc|c|}
\hline
     Quantum & \multicolumn{3}{c}{Trend of Complexity $C(N)$} & \multicolumn{2}{|c|}{PODI pair} & Complexity  \\
     System & Decreasing & Convex & Increasing & $\aa$ & $\bb$ & Character \\
\hline \hline
     Atoms & $\aa \geq 9.09 \, \bb$ & $\aa< 9.09 \, \bb$ & $\aa< 3.33 \, \bb$  & 0 & 0.42 &
     $\aa < \bb, \aa=0$  \\
    (norm=4.095)  &  & \& $\aa \geq 3.33 \, \bb$ &  &  &  & Order  \\
    \hline
    Atomic Clusters  & $\aa \geq 2.90 \, \bb$ & $\aa< 2.90 \, \bb$  &
    $\aa< 1.27 \, \bb$ & 10.57 & 1.17 & $\aa > \bb$  \\
    (norm=$2.88 \times 10^{-5}$) &  & \& $\aa \geq 1.27 \, \bb$  &  &  &  & Disorder  \\
    \hline
    Nuclei  & $\aa \geq 2.47 \, \bb$ & $\aa< 2.47 \, \bb$ &
    $\aa< 1.25 \, \bb$ & 10.98 & 0 & $\aa > \bb, \bb=0$  \\
    (norm=$2.48 \times 10^{-5}$) &  & \& $\aa \geq 1.25 \, \bb$  &  &  &  & Disorder  \\
    \hline
    Bosons  & $\aa \geq 250 \, \bb$ & $\aa< 250 \, \bb$      &
    $\aa< 15.39 \, \bb$ & $120$ & $0.44$ & $\aa \gg \bb$  \\
    (norm=$5.6 \times 10^{-3})$ & & \& $\aa \geq 15.39 \, \bb$ & & & & Disorder \\
\hline
\end{tabular}
    \caption{Analysis of results for the PODI pair $(\aa,\bb)$ and
    complexity character of various quantum systems.}\label{tab:table}
\end{table}

Of all four quantum systems,  atoms and bosons are exceptional, in
the following sense.
\begin{itemize}
    \item Atoms: In atoms there are very hard oscillations of
    complexity measures with $N$ ($=Z$, atomic number). This is
    displayed in the corresponding (fourth) plot of Fig.
    \ref{fig:resatoms}, where we employ, just for the sake of
    comparison, the values $\aa=0$ and $\bb=0.42$, originating
    from the fitting $C(N)=\SDL (N)$ for closed shells
    (fifth plot of Fig. \ref{fig:resatoms}).
    It is seen that the overall behavior can not be characterized
    with certainty as monotonically increasing, decreasing or convex. It is clear
    that our proposed two-step procedure to find $(\aa,\bb)$ and
    the complexity character i.e. first to observe the general
    trend of $C(N)$ and then fit $\SDL (N)$ to $C(N)$ is not
    straightforward for atoms, if we insist to employ the dependence
    of $C$ on all the values of the atomic number $N=Z$. Is seems
    that details of our method should be adjusted, to some extent,
    to available data and/or the specific system under
    consideration. \\
    On the other hand, one can find the complexity character based on
    specific characteristics of the system that may simplify the
    whole picture (something that can be useful for further studies of complexity
    especially in classical chaotic systems).
    In the case of the atoms, such a simplification should be
    the choice of the closed shells atoms. Thus, we organize our results
    to define decreasing, convex and increasing behavior using the trend
    of the closed shells atoms ($Z=2,10,18,36,54$). This choice
    for atoms implies an increasing trend for $C(N)$ and a
    optimal pair $(\aa,\bb)=(0,0.42)$. The same trend has been
    observed in a series of papers employing various models and
    methods \cite{PaperCMP,PaperPCMK,Panos08,Boorgoo07,Sanudo08}.
    \item Bosons: The increasing area dominates, while
    a very narrow decreasing area exists for very small values of
    $\bb$. The optimal pair is $(20,0.08)$.
    Extending our calculations in the $\aa-\bb$ plane for $0\leq \aa \leq 120$, and
    $0\leq \bb \leq 120$ (we do not display this extension in Fig. \ref{fig:resboson}),
     we see that the best SDL-LMC fit is attained
    for $\aa=120$ and $\bb=0.44$. Regardless of the exact proper value of $\aa$, which is certainly
    larger than 120, we are sure that $\aa \gg \bb$. This is a clear indication that a bosonic
    system is to a high degree disordered, as expected from the absence
    of the Pauli principle.
\end{itemize}

We observe in Fig. \ref{fig:resatoms}-\ref{fig:resboson} and the
norms of Table \ref{tab:table}, that the curves $C(N)$ and $\SDL
(N)$ almost coincide for the optimal $(\aa,\bb)$ in clusters and
nuclei, while for atoms and bosons, they show a similar pattern.
This is the best result we can obtain.

It is concluded that three out of the four considered quantum
systems (atomic clusters, nuclei and bosons) cannot grow in
complexity or organize themselves as the number of particles
increases, because $C(N)=\GG_{\aa,\bb}(N)$ is a decreasing
function of $N$. The case of atoms is different, as discussed
above. The question is open whether those systems can show
self-organization (organized complexity), when they form more
complicated structures. At the moment, our indicative result is
that for atoms complexity increases with the atomic number. This
is interesting and promising for future research, because nature
employs atoms as the basic building blocks of molecules and
macromolecules related to biology. The next step is obviously to
change their environment by influencing them by an external
factor. In \cite{FerezDehesa.Paper} it was found that increasing
the magnetic field applied on excited electrons of atoms, the
information content of electrons in specific states is exchanged,
while there is simultaneously an energy level avoided crossing of
the corresponding states. An other idea is to study the effect of
confinement \cite{Sen05}, simulating, in a way, the effect of the
\emph{environment}.

The most obvious objection to SDL and LMC measures of complexity
$\SDL$ and $C$ respectively, is that any function of $S$ e.g.
$\GG_{\aa,\bb}(\Delta)$, $(\Delta=S/S_{\rm max})$ does not give
anything new, compared with $S$. It is still a function of $S$.
This might be true for some cases, but in our specific treatment
of quantum many-body systems with $N$ particles, the function
$S(N)$ is different than $S_{\rm max}(N)$, providing a non-trivial
dependence of $\GG_{\aa,\bb}(\Delta(N))$ on $N$.

%%%%%%%%%%%%%%%%%%%%%%%%%%%%%%%%%%%%%%%%%%%%%%%%%%%%%%%%%%%
%%%%%%%%%%%%%%%%%%%%%%%%%%%%%%%%%%%%%%%%%%%%%%%%%%%%%%%%%%%

%%%%%%%%%%%%%%%%%%%%%%%%%%%%%%%%%%%%%%%%%%%%%%%%%%%%%%%%%%%
%%%%%%%%%%%%%%%%%%%%%%%%%%%%%%%%%%%%%%%%%%%%%%%%%%%%%%%%%%%

%\section{Numerical Results}\label{sec:Numerical}

\section{Conclusions}\label{sec:Conclusions}

It is seen from Table \ref{tab:table} that for bosonic systems
$\aa$ is very large and $\bb$ very small, $\aa\gg \bb$, which
implies that they are \emph{disordered}, as expected intuitively,
because of the absence of the Pauli exclusion principle. Atoms can
be considered as \emph{ordered} $(\aa=0)$. This seems reasonable
and might be expected for a system created by filling with
electrons, one by one, well defined orbitals in the Coulomb field
of the nucleus. Complexity measures are oscillating hard in atoms,
but our experience leads us to the realization that the proper way
to assess complexity is to define the overall trend by observing
the closed shells atoms. A similar conclusion has been drawn for
atoms in \cite{PaperPCMK} and recently in \cite{Panos08}, where
$\aa=0.085$ and $\bb=1.015$, obtained with discrete (fractional)
occupation probabilities of atomic orbitals, indicating an ordered
system. In the same spirit, nuclei and atomic clusters are (less)
\emph{disordered} systems $(\aa>\bb)$ lying between atoms and
bosons. Thus, a character of order or disorder can be assigned to
each system.

\begin{figure}[hb]
 \centering
    \includegraphics[height=7.0cm,width=7.0cm]{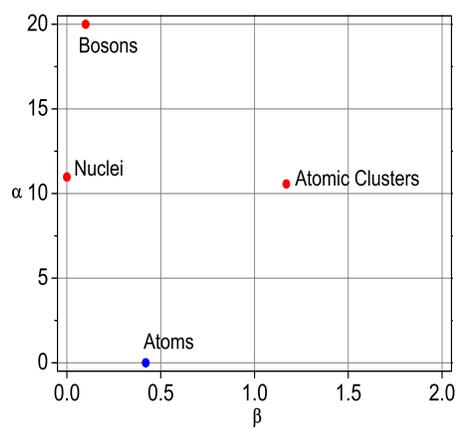}
    \caption{PODI values $(\aa,\bb)$ for atoms, atomic clusters, nuclei and bosons.
    The $\aa$-axis corresponds to "disorder" while
    the $\bb$-axis to "order". \label{fig:abtotal}}
\end{figure}

In other words, a well specified PODI categorization emerges
naturally. Specifically, the inequality $\aa > \bb$ implies a
character of \emph{disorder}, while if $\aa <\bb$ the system has
the character of \emph{order} and there is the possibility of
gradual change of their character from order to disorder realized
in the chain Atoms-Atomic Clusters-Nuclei-Bosons (from left to
right), or equivalently from top to bottom in Table
\ref{tab:table}. It is seen in Fig.\ref{fig:abtotal} that nuclei
and bosons lie very close to the $\aa$-axis (disorder), atoms on
the $\bb$-axis (order), while for atomic clusters the disorder
character is much stronger than that of order.

The proper values of $(\aa,\bb)$ are obtained by an overall
fitting of $\GG_{\aa,\bb}(N)$ with two free parameters to a single
curve $C(N)$ (no free parameters). Hence our statement that a
quantum system is ordered or disordered comes from an overall
comparison of the corresponding values of $\aa$ and $\bb$, i.e.
for a range of values of $N$. This enables us to find the optimal
pair $(\aa,\bb)$. If one wishes to obtain values $(\aa,\bb)$
imposing the condition $\GG_{\aa,\bb}(N)=C(N)$, for a fixed value
of $N$, then one has to fit a two parameter formula
$\GG_{\aa,\bb}(N)$ to just a single value $C(N)$, obtaining a
dependence $\bb(\aa)$. We also observe that for all systems
studied in this work the functional of disequilibrium $D(N)$ is
very similar to $\SDL(N)$ and $C(N)$. The same observation was
made in \cite{Panos08}. Thus a new measure of complexity emerges.

The usefulness of our approach may be validated pragmatically, by
future applications of PODI to several other quantum and classical
systems. First, it should give reasonable results appealing to
intuition. Second and more important, it is interesting to
investigate cases where, instead of varying just the number of
particles $N$, one might study the effect on complexity of other
relevant quantities of the systems as well. For example, the next
step may be to study explicitly the effect of correlations between
particles and/or the influence of various theoretical models of
the systems.

It is noted that so far there is no single or perfect quantitative
definition of complexity. It is quite natural that any effort to
provide a specific definition will be met with criticism
\cite{Crutchfield00,Feldman98,Stoop05}. Comments on the validity
of SDL and LMC measures can be found in Sec. 4 of
\cite{PaperPCMK}, where welcome properties of any definition of
complexity are described in detail.

An answer to the question whether complexity shows an increasing
or decreasing trend with $N$, is dictated by the behavior of
$C(N)$. It is seen, that, at least for the models employed in the
present paper, the trend of complexity is clearly decreasing for
atomic clusters, nuclei and bosons, while it is increasing for
atoms. An additional merit of the present work is the proper
choice of the strength of disorder parameter $\aa$ and the
strength of order $\bb$, by imposing the approximate condition
$\SDL(N)=C(N)$ for all values of $N$. This is in accordance with a
relativistic treatment of atoms, where $C(N)$ shows an increasing
behavior as well \cite{Boorgoo07,Sanudo08}. However, in our
opinion, the latter result is not conclusive, because the
corresponding studies are taking into account only the density
distribution in position-space $\rho(\textbf{r})$. A proper,
complete treatment of information entropy $S$ and disequilibrium
$D$ should take into account both position- and momentum-spaces,
indicated by relations $S=S_r+S_k$ and $D=D_r \cdot D_k$, employed
in our present calculations.

So far, $(\aa,\bb)$ indices have been found for quantum systems.
The same PODI procedure can be applied to quantify order versus
disorder in classical systems, provided that they can be treated
probabilistically. Such an application might be the use of
probability densities in chaotic mappings, enabling researchers to
quantify and observe the change of strengths of disorder $\aa$ and
order $\bb$, while non-linear systems evolve through various
phases of chaotic behavior and routes to chaos. Finally, one may,
in principle associate to any probability distribution depending
at least on one or more parameters describing a system, a PODI
pair and a corresponding complexity character, under certain
conditions. This case can be examined in a future work.

The present study focuses on the dependence of complexity on just
one parameter, i.e. the number of particles. The same method might
be extended to assess the effect of other relevant quantities,
which are of interest in the context of a specific level of
description of a system. Another point, related to the latter one,
is that in order to answer a specific question about organized
complexity and the interplay of order versus disorder, one might
need a sophisticated model of the dynamics of the system, or, in
some cases, a simplified one, describing only salient but
necessary features. This might be sufficient for a special aim.
From this point of view, the merit of a model describing a system
will be assessed, not only by its ability to reproduce exactly
experimental results or simulate its behavior accurately, but by
being robust enough to uncertainties or errors in description.
This is facilitated by the fact that an inequality between $\aa,
\bb$ can be satisfied for a wider range of values, instead of an
approximate equality $\aa \simeq \bb$.

In brief, we propose a recipe to find a Pair of Order-Disorder
Indices $(\aa,\bb)$ (PODI) in order to classify the order versus
disorder contribution to complexity of any system (in our case
quantum) described probabilistically. This enables us to assign to
systems the category of order ($\aa <\bb$) or disorder ($\aa
>\bb$) and classify them accordingly.

We conclude that the atom is the only system out of four, studied
here, which is \emph{ordered and growing in complexity with the
number of particles (electrons).} We conjecture that this is an
information-theoretic explanation, supporting the fact that atoms
are used by nature, as suitable components to construct larger
structures i.e. molecules and macromolecules. It is also a first
step, with encouraging results, to fulfill quantitatively
Wheeler's plan as the fundamental theory of the Cosmos.

\acknowledgements K. Ch. Chatzisavvas is supported by a
Post-Doctoral Research Fellowship of the Hellenic State Institute
of Scholarships (IKY).

\clearpage
\newpage

\clearpage
\newpage

\end{document}